# Game-theory-based analysis on interactions among secondary and malicious users in coordinated jamming attack in cognitive radio systems

[1]Ehsan Meamari   [2]Khadijeh Afhamisisi   Hadi Shahriar Shahhoseini

ehsanmeamari@gmail.com   afhami@iust.ac.ir   hshsh@iust.ac.ir

[1, 2,3] Electrical Engineering Department, Iran University of Science and Technology, Tehran, Iran, 1684613114, Tel: +98 21 77240540

**Abstract**: IEEE 802.22 standard utilizes cognitive radio (CR) techniques to allow sharing unused spectrum band. The cognitive radio is vulnerable to various attacks such as jamming attacks. This paper has focused on coordinated jamming attacks. A simple strategy for secondary users is to change their bands and switch to other appropriate bands when the jamming attack is occurred. Also, the malicious users should switch to other bands in order to jam the secondary users. To address this problem, a game theoretical method is proposed to analyze coordinated jamming attacks in CR. Then, using Nash equilibrium on the proposed game, the most appropriate bands have been found to switch as well as the optimal switching probabilities for both secondary and malicious users. Meanwhile, effects of different parameters like the number of malicious users are investigated in changing the optimal switching probabilities by analysis on the model.

## I. Introduction

The IEEE 802.22 standard is a solution to dynamic spectrum access (DSA) that is known as the cognitive radio. To enable DSA networks, the use of cognitive radio technology is being considered due to its ability to rapidly and autonomously adapt operating parameters to change the requirements and conditions.

In cognitive radio, secondary users get access to the licensed spectrum if there is no primary user. The secondary users must leave the spectrum as soon as the primary users need to access the band[1, 2]. Thus, the secondary users always observe the spectrum band in order to avoid collision with the primary users.

Cognitive radio operates on wireless media. The wireless network makes security vulnerabilities unavoidable. An attack on the cognitive radio network can be designed as any kind of activity that leads to unacceptable interference with the licensed primary users or missed opportunities for the secondary users. The malicious users try to prevent the secondary or primary users using the spectrum band. Numerous papers have concentrated on studying the behavior of malicious users in IEEE 802.22[3-5]. There are three types of attacks in physical layer for the secondary users: Primary User Emulation Attacks (PUE)[6, 7], Reporting False Sensing Data Attacks (RFSD)[8, 9] and Jamming Attacks[10, 11].

In the distributed spectrum sensing, malicious users can send false data to the fusion center which leads to incorrect spectrum sensing and decision. This attack is called Reporting False Sensing Data Attack (RFSD). Authors[12] propose a new method which employs a variable number of samples instead of fixed samples for sensing data. Furthermore, the RFSD attacks have studied[13] when the number of malicious users is unknown and an onion-peeling approach is proposed to defend against multiple untrustworthy secondary nodes.

Malicious users can modify air interface to minimize signal characteristics of the primary users. Thus the secondary users deem it as the primary user's signal and leave the band[14]. This attack was first introduced by Chan and Pank[15]. The first analytical model to confront PUE attacks based on the received power of primary and malicious users is studied by Anand[16]. A Bayesian game is considered[17] to model PUE attacks according to the power received from malicious and secondary users.

Li and Han studied a PUE attack by a dogfight game in cognitive radio networks[18]. They argued that in PUE attack the secondary users should switch to appropriate bands like jamming attack. So they solved the PUE attack by a game theoretical approach and determined the appropriate bands and optimal switching probabilities for both secondary and malicious users. Then, they introduced a Partially Observable Markov Decision Process (POMDP) model to analyze the PUE attack. They presented a model which is only utilized for single malicious user, and further improved their work[19].

The jamming attack is harmful in these networks since the malicious users wish to interrupt communication sessions of the secondary users by making interference. When the generated interference of the malicious users is high enough, they can substantially decrease the communication performance or even terminate it by running DoS attacks.

The jamming attack in multichannel cognitive radio networks is introduced[9], which is aimed to avoid the jamming signal through frequency hopping. A Markov Decision Process (MDP) model for jamming attacks is introduced[20] as a countermeasure with a learning method being suggested for the secondary users are able to learn the access pattern of the primary users as well as the number of malicious users. This paper extends the previous work done[21] and recommends a stochastic game framework for anti-jamming defense design, which can accommodate dynamic spectrum opportunity and channel quality, as well as the strategy of secondary and malicious users.

A simple strategy of secondary and malicious users for defense and attack in jamming attacks through CRN is switching to another frequency band, but determination of the appropriate bands for switching should not be overlooked. Meanwhile, it is necessary to determine optimal probabilities of switching to appropriate bands. A cooperated jamming attack is studied by Tan, Sengupt and Subbalakshmi[22]. They determined appropriate bands for switching in addition to optimal switching probabilities just for the malicious users. One major drawback of their work[22] is that it does not determine appropriate bands and optimal switching probabilities for the secondary users. Another drawback of their work[22] is its inability to consider the role of primary users. But in this contribution, a game theoretical method has been utilized for analysis of a coordinated attack by considering the role of secondary, malicious and primary users followed by deriving optimal bands for the strategies of both malicious and secondary users.

In this model, there is a base station for secondary users that is called secondary-BS. The secondary-BS may be a fusion center (FC) or a typical secondary user. Also, a base station for malicious users is used which is known as the malicious-BS (Fig. 1). The malicious-BS manages malicious users and sends them to an appropriate band where no other malicious user exists. Also the secondary-BS manages secondary users and informs them about the free bands.

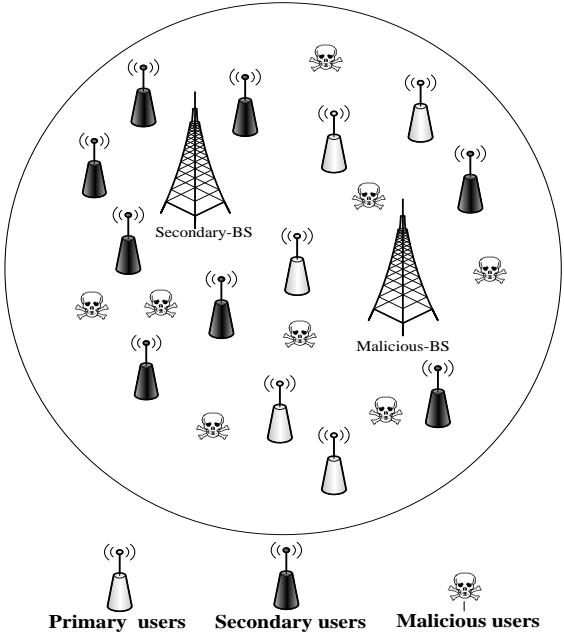

Fig. 1. The malicious and the secondary users with their base stations

In this article, it is supposed that secondary users make use of the switching strategy against jamming attack. The main purpose here is to determine appropriate bands for switching as well as optimal switching probabilities for both secondary and malicious users. Thus, the spectrum band is classified in accordance with presence or absence of primary, secondary and malicious networks. Thereby, appropriate bands were found for the next time slot of secondary and malicious users based on the proposed game. Then the coordinated attack was formulated as a cooperated game regarding the role of secondary, malicious and primary users while the optimal switching probabilities was calculated for both secondary and malicious users via Nash equilibrium of the proposed game.

A reward value was defined for successful transmissions of secondary users and a reward value for successful attacks of malicious users in addition to cost values for switching to other bands for both of them. It was demonstrated that total payoffs of secondary and malicious users, depend just on their own cost and reward values, rather than cost and reward values of their rival. But their optimal switching probabilities of secondary and the malicious users are dependent on cost and reward values of the rival instead of their own cost and reward values.

The rest of this paper is organized as follows: System model, assumptions and proposed classification are described in section II. In section III, a coordinated game is formulated between malicious and secondary users which has derived an analytical expression for payoffs and optimal switching probabilities of them. Numerical results are discussed in section IV. Finally, summary and conclusions are made in section V.

## II. System model

The IEEE 802.22 standard defines a point-multipoint air-interface, composed of a base station and several Consumer Premise Equipment (CPEs). The base station is responsible for collecting spectrum sensing information provided by several CPEs and controlling the medium access.

Suppose a network with $n_N$ spectrum bands which can be occupied with $n_p$ primary users and $n_s$ secondary users. The secondary users are managed with a central base station (secondary-BS) to get the optimal performance. It is supposed that there are $n_m$ malicious nodes managed by central base station (Malicious-BS) to establish a coordinated attack on the secondary users. The primary users can occupy the band at any time and we further assume that the malicious users do not want to establish the jamming attack on the primary users like some papers[20, 21]. Fig .1 demonstrates the secondary and malicious users along with their base stations.

Time is divided into several time slots. At the beginning of each time slot, both secondary and malicious users are synchronized with primary users. The secondary user can access the band if no primary activity is sensed after receiving the signal. The malicious user senses the secondary signal when there no primary signal is detected and then tries to jam the band if the malicious user finds the secondary user without any primary users. Fig. 2 depicts priority of the sensing spectrum band for secondary, primary and malicious users.

The spectrum occupancy of each channel for primary users is modeled by a two-state Markov chain (ON-OFF model) which is illustrated in Fig. 3.

Two states of this Markov chain are idle ($I$, no primary user exists in the band) and busy ($B$, one primary user exists in the band). The spectrum occupancy hops between these two states depending on the value of $\alpha$ and $\beta$. For simplicity, it is assumed that $\alpha$ and $\beta$ have the same values for all $n_N$ spectrum bands.

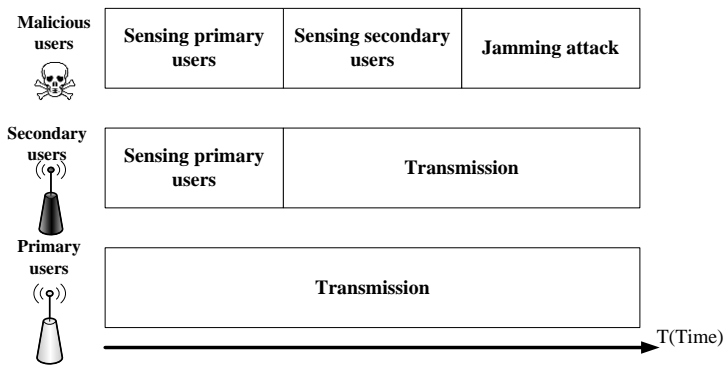

Fig. 2. The priority of spectrum band Sensing for the primary, secondary and malicious users

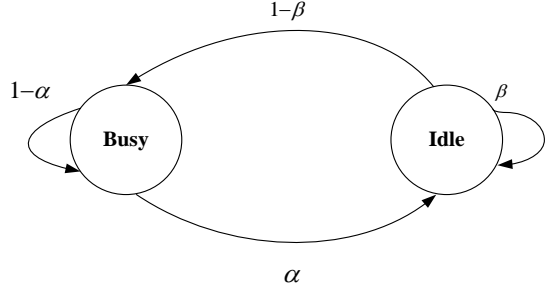

Fig. 3. Discrete-Time Markov Chain (DTMC) model for spectrum occupancy of each channel for the primary users[23]

By solving the steady-state Markov chains in Fig. 3, the probabilities for a specific channel being busy by one primary user ($P_{pB}$) or idle ($P_{pI}$) in an arbitrary time slot are given by Eq. (1) and Eq. (2), respectively:

$$P_{pB} = \frac{1-\beta}{1-\beta+\alpha} \qquad (1)$$

$$P_{pI} = \frac{\alpha}{1-\beta+\alpha} \qquad (2)$$

It is also assumed that $\alpha$ and $\beta$ have the same values for all $n_N$ spectrum bands like some works[20, 21]. So $P_{pI}$ and $P_{pB}$ show the same values for all $n_N$ spectrum bands.

In the proposed model, there are three kinds of players: primary, secondary and malicious users. Thus, taking into account the absence or presence of each type of these users, eight different states can be defined for occupation of bands (Fig. 4a) as below:

1- The bands are occupied by malicious and secondary users.
2- The bands are occupied by secondary users.
3- The bands are occupied by malicious users.
4- The bands are occupied by secondary, malicious and primary users.
5- The bands are occupied by primary and malicious users.
6- The bands are occupied by secondary and primary users.
7- The bands are occupied by primary users.
8- Free bands

In our model, it is supposed that malicious and secondary users send information of their own occupied bands to their base stations. According to the current situation of the next time slot, this information is useful for the base station to decide what the distribution of its members is. Each one of secondary and malicious users has the mere information about the state of its own band and sends it to the base station. That is why malicious-BS and secondary-BS represent different information about occupation of the bands. Fig. 4b and Fig. 4c present different states of the bands from the view point of secondary-BS and malicious-BS, respectively. According to Fig. 4b, four classes are seen from the view point of the secondary-BS for the states of bands ($a, b, c$ and $d$):

1- The bands are occupied by malicious and secondary users ($a$).
2- The bands are occupied by secondary users ($b$).
3- The bands are occupied by secondary and primary users ($c$).
4- The bands with unknown states ($d$).

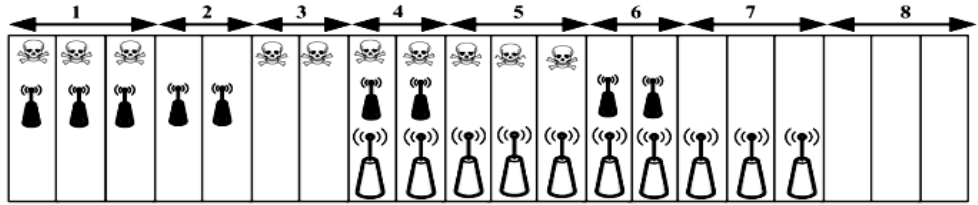
Fig. 4a. The eight states of bands by presenting the malicious, primary and secondary users

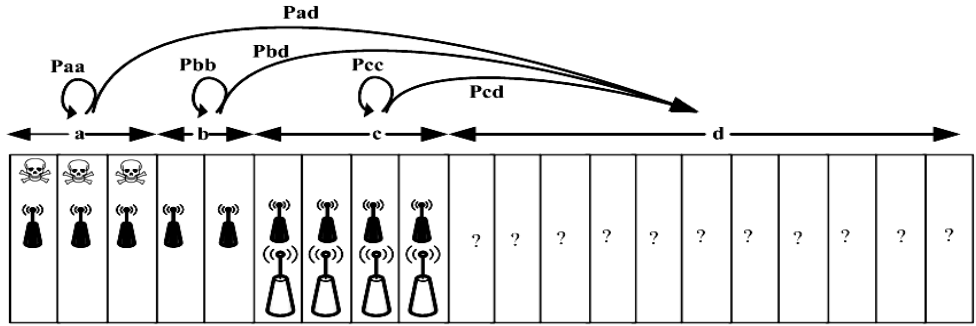
Fig. 4b. Four states of bands from the viewpoint of the secondary-BS

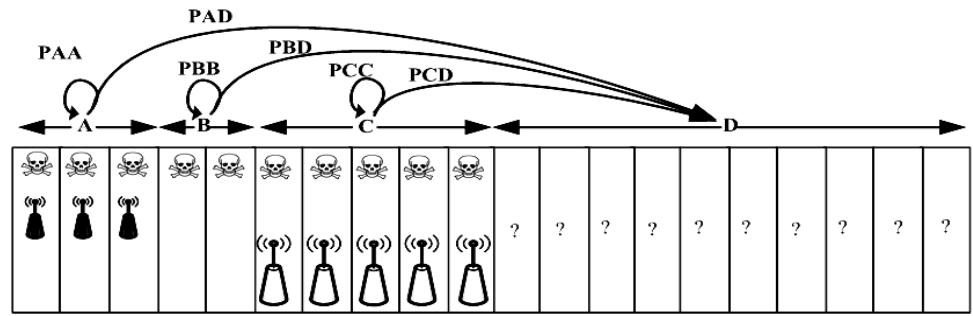
Fig. 4c. Four states of bands from the viewpoint of the malicious-BS

In Fig. 4a, the secondary users in states 4 and 6 sense the primary users, thus they cannot use these bands and the malicious users do not jam the secondary users. This leads to a situation in which the secondary users cannot recognize presence of the malicious users in the bands of states 4 and 6 and they would report incomplete information to the secondary-BS. Thus, the secondary-BS cannot distinguish between states 4 and 6 and behaves it as if it is just state $c$. There is no secondary user in the bands of states 3, 5, 7 and 8. Therefore, the secondary-BS does not have any information about these states. As a result, the states 3, 5, 7 and 8 look the same from the viewpoint of secondary-BS with state $d$ (Fig. 4b). Similarly, according to Fig. 4c, there are four classes of states ($A, B, C$ and $D$) from the viewpoint of malicious-BS:

1- The bands are occupied by malicious and secondary users ($A$).
2- The bands are occupied by malicious users ($B$).
3- The bands are occupied by malicious and primary users ($C$).
4- The bands with unknown state ($D$).

The malicious-BS deems states 4 and 5 as state $C$ while observing states 2, 6, 7 and 8 as state $D$.

# III. Nash equilibrium

This section models attack and defense problems using a static game. Goal of the game is to find appropriate bands for switching as well as optimal switching probabilities to maximize secondary and malicious users' payoff functions simultaneously.

Fig. 4b shows that secondary users of the bands of state $a$ cannot switch to the bands of states $b$ and $c$, since secondary-BS knows that other secondary users exist in the bands of states $b$ and $c$ and tries to avoid collision between them. Therefore, there are only two strategies for secondary users in the bands of state $a$: staying in the bands of state $a$ or switching to the bands of state $d$. The probability of staying in the bands of state $a$ is denoted with $P_{aa}$ while the probability of switching from the bands of state $a$ to those of state $d$ is represented with $P_{ad}$ as shown in Fig. 4b.

As described before, from the viewpoint of malicious-BS and secondary-BS, the spectrum bands are classified into four states. Moreover, there are only two different strategies for each state. The strategies in each state are depicted in Figs. 4b and 4c for secondary and malicious users, respectively. Eq. (3) and Eq. (4) address the same strategies for other states:

$$P_{ll} = 1 - P_{ld} \quad l = a,b,c \quad (3)$$

$$P_{kk} = 1 - P_{kD} \quad k = A,B,C \quad (4)$$

Having defined the appropriate bands for switching, secondary and malicious users should now calculate the optimal probabilities for either staying in their band or switching to other appropriate bands. A game theory is used to find these optimal switching probabilities. Therefore, the payoff functions of secondary and malicious users should be maximized in order to find Nash equilibrium.

### A. General algorithm for calculating Nash equilibrium:

Both malicious-BS and secondary-BS should calculate the payoff function in every state. Secondary and malicious users have only two strategies: staying in their band or switching to other bands (state $d$ for the secondary users or state $D$ for the malicious users). We define a payoff function for those strategies:

The secondary users' payoff function for staying in the band $= U_{ll}$, $l = a,b,c$

The malicious users' payoff function for staying in the band $= U_{kk}$, $k = A,B,C$ (5)

The secondary users' payoff function for switching $= U_{ld}$, $l = a,b,c$

The malicious users' payoff function for switching $= U_{kD}$, $k = A,B,C$ (6)

In order to calculate the payoff functions, the cost value of switching to a band is shown with $C_m$ for malicious users and $C_s$ for secondary users. The value of malicious reward by performing a successful jamming attack on secondary users is $G_m$ while the value of secondary reward by establishment of a successful connection without malicious and primary users is denoted by $G_s$. Penalty of secondary user, attacked by the malicious users is $L_s$. As current connections are interrupted, they call for an extra effort to re-establish a new connection.

The secondary users can establish a connection and obtain a reward $G_s$ in a band when there are no primary, secondary or malicious users in their band like the bands of state 2 in Fig. 4a. $E_{js}$ represents the number of bands expected by secondary-BS and malicious-BS to occupy just with secondary users in the next time slot.

Malicious users can jam a band and obtain the reward $G_m$ when there is only one secondary user in its band at the absence of primary users like the bands of state 1 in Fig. 4a. In this situation, the loss of secondary user would be given by $L_s$. $E_{jsm}$ is the number of bands that secondary-BS and malicious-BS expect to be occupied only by secondary and malicious users in the next time slot.

Other states of arrangement for secondary, malicious and primary users in Fig. 4a do not bear any gains or penalties for secondary and malicious users and even no effect on calculating the payoff function of secondary and malicious users.

For calculation of $E_{js}$, the probability of occupying each band just by secondary users should be calculated ($P_{js}$). Moreover, for calculation of $E_{jsm}$, it is necessary to compute the probability of occupation for each band just by secondary and malicious users ($P_{jsm}$). In order to calculate $P_{js}$ and $P_{jsm}$, the probability of occupation of each band by primary, malicious and secondary users should be computed first and these probabilities are specified with $P_p$, $P_m$ and $P_s$, respectively.

Calculations in part (B) and (C) of this section demonstrate if these payoff functions are arranged by the switching probabilities, the payoff functions of secondary (malicious) users at each state can be separated into product of their switching probabilities and a linear function of the switching probabilities for malicious (secondary) users. All of these payoff functions should be maximized together to reproduce Nash equilibrium for this problem. Thus, derivative of payoff ratios for the switching probabilities are taken to simultaneously maximize the payoff functions of malicious and secondary users. This leads to two systems of three equations and three unknowns. The optimal switching probabilities for the secondary users are obtained by solving these two systems.

One system of three equations and three unknowns is based on the switching probabilities of secondary (malicious) users. This system is the result of maximizing the payoff functions of malicious (secondary) users.

Therefore, secondary-BS (malicious-BS) should calculate the payoff functions of malicious (secondary) users while maximizing them to address the system of three equations and three unknowns based on the switching probabilities of secondary (malicious) users. Then, secondary-BS (malicious-BS) will calculate the switching probabilities of secondary (malicious) users by solving the system of equations. Fig. 5 illustrates this algorithm for calculating the optimal switching probabilities.

Table 1. The probability of occupation of one band in state $k$ with the secondary, malicious and primary users

| State | The occupation probability of one band in state $k$ with primary users | The occupation probability of one band in state $k$ with a secondary user | The occupation probability of one band in state $k$ with a malicious user |
|---|---|---|---|
| A | $P_p^A = 1-\beta$ | $P_s^A = P_{aa}$ | $P_m^A = P_{AA}$ |
| B | $P_p^B = 1-\beta$ | $P_s^A = \dfrac{n_s^d}{n_B} \times \dfrac{n_3}{n_d}$ | $P_m^B = P_{BB}$ |
| C | $P_p^C = 1-\alpha$ | $P_s^C = \dfrac{(\dfrac{n_5}{n_d} \times n_s^d) \times (n_4 \times P_{cc})}{n_C}$ | $P_m^C = P_{CC}$ |
| D | $P_p^D = \dfrac{(n_6+n_7)\times(1-\alpha)+(n_8+n_2)\times(1-\beta)}{n_D}$ | $P_s^D = \dfrac{n_b \times P_{bb} + (\dfrac{n_7+n_8}{n_d} \times n_s^d)}{n_D}$ | $P_m^D = \dfrac{n_m^D}{n_D}$ |

## B. optimal switching probabilities for secondary users

Like the algorithm proposed in Fig. 5, $P_p^k$, $P_m^k$ and $P_s^k$ (the occupation probabilities of one band in state $k(k = A,B,C,D)$ by primary, malicious and secondary users, respectively) are derived first (Table 1).

In Table 1, $n_k$ denotes the number of bands at state $k(k = A,B,C,D)$ (Fig. 4c) while $n_i$ gives the number of bands in state $i(i = 1,2,3,4,5,6,7,8)$ (Fig. 4a). Also, $n_m^D$ is the number of malicious users that will switch to the bands of state $D$ in the next time slot (Eq. (7)).

$$n_m^D = n_A \times P_{AD} + n_B \times P_{BD} + n_C \times P_{CD} \tag{7}$$

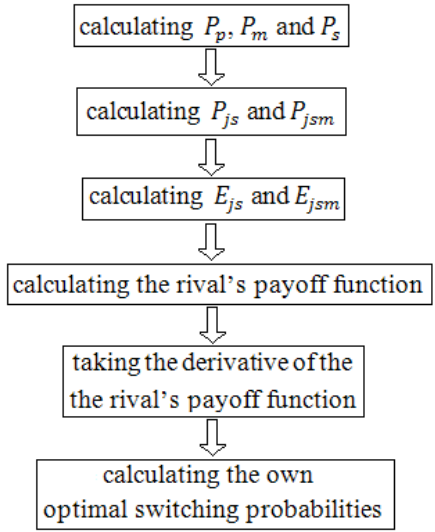

Fig. 5. An algorithm for calculate the optimal switching probabilities

The secondary-BS needs to discover the number of bands at each state of $i(i = 1,2,3,4,5,6,7,8)$ and $k(k = A,B,C,D)$ in order to derive the probability values from Table 1. Since some of this information is unknown, one has to assess the number of bands at all states. Using the information gathered from secondary users, secondary-BS knows the number of secondary users in band $n_a$ and $n_b$. Therefore, $n_1$ and $n_2$ are calculated from Eq. (8).

$$n_1 = n_a, n_2 = n_b \qquad (8)$$

Furthermore, secondary-BS knows that $n_m - n_a$ malicious users are at the states $c$ and $d$. Although secondary-BS does not know exactly how many of these $n_m - n_a$ malicious users are in state $c$ and how many of them are in state $d$. Thus, secondary-BS assumes that malicious users are distributed uniformly through the bands of state $c$ and $d$. Therefore, $n_4$ and $n_6$ are derivable as demonstrated by Eq. (9):

$$n_4 = n_c \times (\frac{n_m - n_a}{n_c + n_d}), \ n_6 = n_c - n_4 \qquad (9)$$

So, $n_m - n_1 - n_4$ malicious users are in state $d$ and then number of the primary users ($n_p$) is:

$$n_p = n_N \times P_{pB} \qquad (10)$$

Also, there are $n_p - n_c$ primary users in the bands of state $d$. It is trivial to find $n_5$, $n_3$, $n_7$ and $n_8$, respectively:

$$n_5 = \frac{(n_p - n_c) \times (n_m - n_1 - n_4)}{n_d}$$
$$n_3 = n_m - n_4 - n_5 - n_1$$
$$n_7 = n_p - n_4 - n_5 - n_6$$
$$n_8 = n_d - n_3 - n_5 - n_7$$
$$n_A = n_a$$
$$n_B = n_3$$
$$n_C = n_4 + n_5$$
$$n_D = n_N - n_A - n_B - n_C \qquad (11)$$

We suppose that $P_{jsm}^k$ is the probability that one band in state $k$ to be occupied just by secondary and malicious users at the absence of primary users. This probability can be calculated according to Eq. (12).

$$P_{jsm}^k = P_s^k \times P_m^k \times (1 - P_p^k) \quad k = A, B, C, D \qquad (12)$$

For the bands of state $k$, $E_{jsm}^k$ is the number of bands that secondary-BS expects to be occupied just by secondary and malicious users without primary users in the next time slot.

$$E_{jsm}^k = n_k \times P_{jsm}^k \quad k = A, B, C, D \qquad (13)$$

The payoff functions for malicious users that remain in their band can be computed from Eq. (14):

$$U_{kk} = G_m \times E_{jsm}^k \quad k = A, B, C \tag{14}$$

$U_{AA}$, $U_{BB}$ and $U_{CC}$ are calculated from Eq. (14). The payoff functions for malicious users that switch to the bands of state $D$ are:

$$U_{kD} = \frac{n_k \times P_{kD}}{n_m^D} \times (G_m \times E_{jsm}^D) - C_m \times n_k \times P_{kD} \tag{15}$$
$$k = A, B, C$$

Therefore, $U_{AD}$, $U_{BD}$ and $U_{CD}$ are obtained from Eq. (15). Nash equilibrium of this game occurs where the payoffs of two strategies (stay and switch) are equal.

$$\begin{aligned} U_{AA} &= U_{AD} \\ U_{BB} &= U_{BD} \\ U_{CC} &= U_{CD} \end{aligned} \tag{16}$$

If equalizations of Eq. (16) are assigned, three linear functions for the switching probabilities of secondary users can be obtained.

$$\begin{aligned} F(P_{ad}, P_{bd}, P_{cd}) &= 0 \\ G(P_{ad}, P_{bd}, P_{cd}) &= 0 \\ H(P_{ad}, P_{bd}, P_{cd}) &= 0 \end{aligned} \tag{17}$$

All of three equalizations should be established simultaneously to determine Nash equilibrium of this problem. This problem leads to a system of three equations and three unknowns. By solving this system, the optimal switching probabilities for secondary users will be obtained.

At each time slot, every secondary user senses its band and reports the secondary-BS state of its band. The secondary-BS, calculates the number of bands in each state with Eqs.(8) to (11) according to its gathered information. Afterwards, the secondary-BS computes the payoff functions of malicious users and optimal switching probabilities for secondary users. Thus, according to the optimal switching probabilities derived, secondary-BS decides which secondary user should stay at their bands and which secondary user should switch to the bands of state $d$. The secondary-BS should organize the switching of secondary users in the bands of state $d$ in order to avoid their collision.

### C. optimal switching probabilities for malicious users

$P_p^l$, $P_m^l$ and $P_s^l$ are the probabilities of occupation for a band in state $l (l = a,b,c,d)$ by primary, malicious and secondary users, respectively. These probabilities are computed in Table 2.

In Table 2, $n_l$ is the number of bands in state $l$, while $n_s^d$ gives the number of secondary users that will switch to the bands of state $d$ at the next time slot (Eq. (18)).

$$n_s^d = n_a \times P_{ad} + n_b \times P_{bd} + n_c \times P_{cd} \tag{18}$$

For calculating the probability values from Table 2, malicious-BS needs to know the number of bands at each state of $i(i = 1,2,3,4,5,6,7,8)$ and $l(l = a,b,c,d)$.

Since it does not know some of them, one has to just assess them. Similar to part (B), malicious-BS assesses the bands of different states. Table 3 lists the assessments on the number of bands by malicious-BS. For calculation of the number of different states, different arrays of Table 3 should be derived sequentially.

Table 2. The probability of occupation one band in state $l$ with secondary, malicious and primary users

| State | The occupation probability of one band in state $l$ with a primary user | The occupation probability of one band in state $l$ with a secondary user | The occupation probability of one band in state $l$ with a malicious user |
|---|---|---|---|
| a | $P_p^a = 1 - \beta$ | $P_s^a = P_{aa}$ | $P_m^a = P_{AA}$ |
| b | $P_p^b = 1 - \beta$ | $P_s^b = P_{bb}$ | $P_m^b = \frac{n_2}{n_D} \times \frac{n_m^D}{n_b}$ |
| c | $P_p^c = 1 - \alpha$ | $P_s^c = P_{cc}$ | $P_m^c = \frac{(\frac{n_6}{n_D}) + (n_4 \times P_{CC})}{n_c}$ |
| d | $P_p^d = \frac{(n_5 + n_7) \times (1-\alpha) + (n_8 + n_3) \times (1-\beta)}{n_D}$ | $P_s^d = \frac{n_s^d}{n_d}$ | $P_m^d = \frac{n_B \times P_{BB} + (\frac{n_7 + n_8}{n_D} \times n_m^D) + (n_5 \times P_{CC})}{n_d}$ |

It is supposed that $P_{js}^l$ is the probability that one band in state $l$ is occupied just with secondary user and it is computed by Eq. (19).

$$P_{js}^l = P_s^l \times (1 - P_m^l - P_p^l + (P_m^l \times P_p^l)) \quad l = a,b,c,d \quad (19)$$

Thus for the bands of state $l$, $E_{js}^l$ is the number of bands that the malicious-BS expects to be occupied just with secondary users at the next time slot, consequently:

$$E_{js}^l = n_l \times P_{js}^l \quad l = a,b,c,d \quad (20)$$

$P_{jsm}^l$ is the probability of one band in state $l$ being occupied just by secondary and malicious users in the absence of primary users. This probability is calculated from Eq. (21).

$$P_{jsm}^l = P_s^l \times P_m^l \times (1 - P_p^l) \quad l = a,b,c,d \quad (21)$$

Thus, for the bands of state $l$, $E_{jsm}^l$ denotes the number of bands that malicious-BS expects to be occupied just by secondary and malicious users with no primary users at the next time slot.

$$E_{jsm}^l = n_l \times P_{jsm}^l \quad l = a,b,c,d \quad (22)$$

The payoff functions for secondary users which remain in their band are presented in Eq. (23) below:

$$U_{ll} = G_s \times E_{js}^l - L_s \times E_{jsm}^l \quad l = a,b,c \quad (23)$$

Table 3. Assessment of the number of bands in each state by the Malicious-BS

| |
|---|
| $n_1 = n_A$ |
| $n_3 = n_B$ |
| $n_4 = n_C \times (\dfrac{n_s - n_A}{n_C + n_D})$ |
| $n_5 = n_C - n_4$ |
| $n_6 = \dfrac{(n_p - n_C) \times (n_s - n_A - n_4)}{n_D}$ |
| $n_7 = n_p - n_4 - n_5 - n_6$ |
| $n_2 = n_s - n_4 - n_6 - n_1$ |
| $n_8 = n_D - n_2 - n_6 - n_7$ |
| $n_a = n_A$ |
| $n_b = n_2$ |
| $n_c = n_4 + n_6$ |
| $n_d = n_N - n_a - n_b - n_c$ |

Values of $U_{aa}$, $U_{bb}$ and $U_{cc}$ can be derived from Eq. (24). The payoff functions for secondary users that switches to the bands of state $d$ are introduced below:

$$U_{ld} = \frac{n_l \times P_{ld}}{n_s^d} \times (G_s \times E_{js}^d - L_s \times E_{jsm}^d) - C_s \times n_l \times P_{ld} \quad (24)$$

$$l = a, b, c$$

Then, $U_{ad}$, $U_{bd}$ and $U_{cd}$ can be determined from Eq. (24). Such as before, Nash equilibrium of this game for malicious users is where payoffs of two strategies (stay and switch) are equal.

$$\begin{aligned} U_{aa} &= U_{ad} \\ U_{bb} &= U_{bd} \\ U_{cc} &= U_{cd} \end{aligned} \quad (25)$$

By sorting Eq. (25), three linear functions for the switching probabilities of malicious users can be generated.

$$\begin{aligned} f(P_{AD}, P_{BD}, P_{CD}) &= 0 \\ g(P_{AD}, P_{BD}, P_{CD}) &= 0 \\ h(P_{AD}, P_{BD}, P_{CD}) &= 0 \end{aligned} \quad (26)$$

By solving another system of three equations and three unknowns, the optimal switching probabilities for malicious users will be obtained.

At every time slot, each malicious user reports malicious-BS state of its band. The malicious-BS, calculates the number of bands in each state according to the received information (Table 3). Afterwards, malicious-BS adopts to extract the payoff functions of secondary users as well as the optimal switching probabilities of malicious users. Then, according to the computed optimal switching probabilities, malicious-BS decides which malicious users should stay in their bands and which should switch to the bands of state $D$. For avoiding collision between malicious users, malicious-BS should organize their switching in the bands of state $D$.

## IV. Numerical results

In this section, numerical simulation is launched to evaluate performance of the proposed game against coordinated jamming attacks. The total payoff for secondary users is shown with $U_s$ while the total payoff for malicious users is represented by $U_m$.

$$U_s = U_{aa} + U_{ad} + U_{bb} + U_{bd} + U_{cc} + U_{cd}$$
$$U_m = U_{AA} + U_{AD} + U_{BB} + U_{BD} + U_{CC} + U_{CD} \quad (27)$$

Before studying the numerical results, it is better to notice where the total payoff for secondary users ($U_s$) are negative. It means that by using Nash equilibrium for this situation, secondary users will not have any positive payoff from switching to other bands or staying in their bands, so they will leave network for the next time slot, although they may go back to the network after ending the next time slot. The initial values of different analytical results are: $n_N = 50$, $n_s = 10$, $n_m = 10$, $C_S = 1$, $C_m = 1$, $G_s = 50$, $G_m = 75$, $L_s = 100$, $\alpha = 0.4$, $\beta = 0.6$.

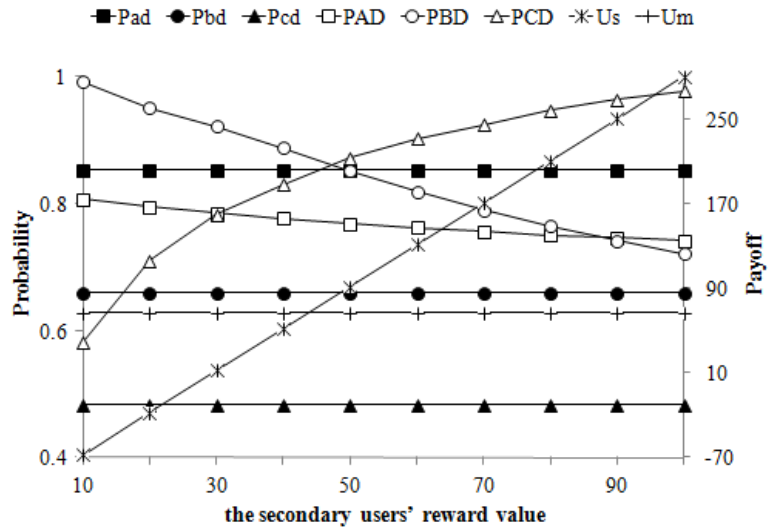

Fig. 6. Payoffs and switching probabilities versus the secondary users' reward value

Fig. 6 demonstrates that increasing $G_s$ does not significantly affect the switching probabilities of secondary users but increases their payoff. The negative payoffs for secondary users in Fig. 6 show that switching strategy for secondary users to avoid jamming attacks involves no payoff for secondary users. Thus, secondary users will leave the network for the coming time slot. Raising $G_s$ does not have any effect on the payoff of malicious users but it tends to change their switching probabilities. With increasing $G_s$ the incentive of malicious users in states $A$ and $B$ are decreased for switching to state $D$ while the incentive of malicious users in state $C$ is increased for switching to state $D$.

Fig. 7 demonstrates that increasing $G_m$ does not affect the payoff of secondary users much but alters their switching probabilities. So the motivation of secondary users in state $b$ is decreased for switching to state $d$ and the motivation of secondary users in bands $a$ and $c$ is increased for switching to state $d$. Increasing $G_m$ does not affect the switching probabilities of malicious users but increases their payoff instead.

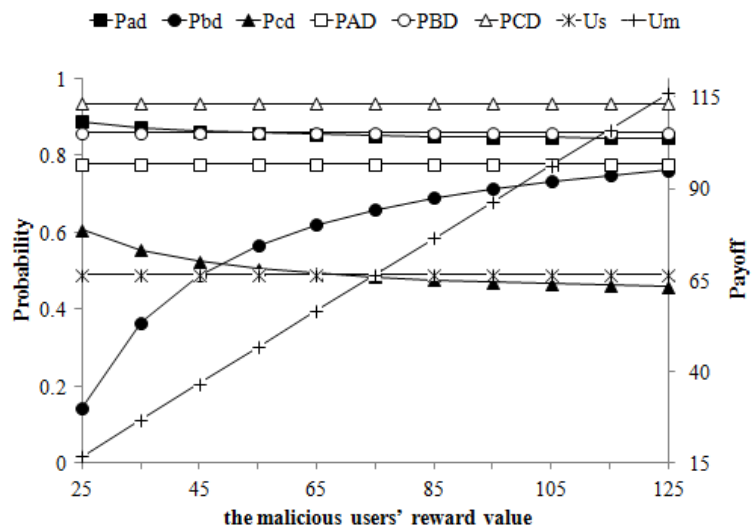

Fig. 7. Payoffs and switching probabilities versus the malicious users' reward value

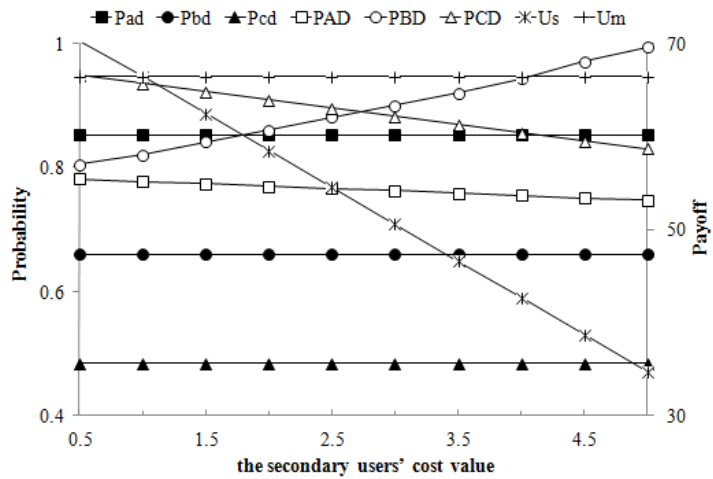

Fig. 8. Payoffs and switching probabilities versus the secondary users' cost value

Fig. 8 shows that increasing $C_s$ does not change the secondary users' switching probabilities but decreases their payoff. Also, increasing $C_s$ does not change the malicious users' payoff but causes the malicious users in state $B$ to have more motivation to switch to state $D$ and decreases the probability of switching them from state $A$ and $C$ to state $D$.

The effect of changing $C_m$ is demonstrated in Fig. 9. The results demonstrate that increasing $C_m$ does not considerably affect the payoff of secondary users but changes their optimal switching probabilities. Decreasing the motivation of secondary users in state $b$ for switching to state $d$ and increasing the motivation of secondary users in states $a$ and $c$ for switching to state $d$ are consequences of increasing $C_m$. Also increasing $C_m$ will decrease the payoff of malicious users but will not alter their switching probabilities much.

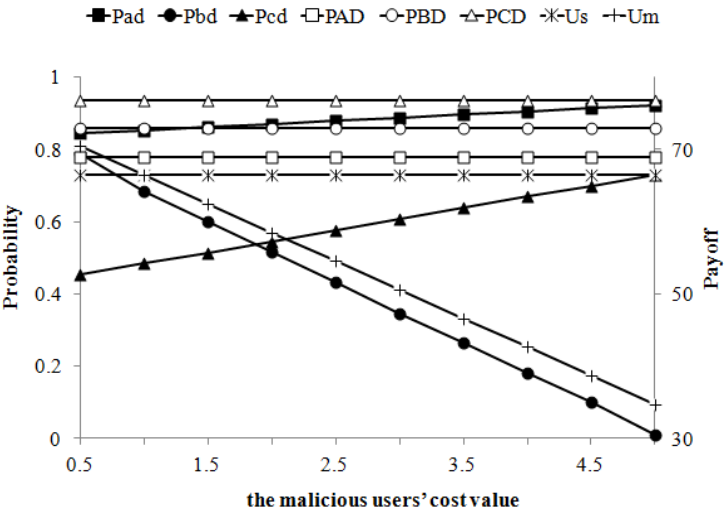

Fig. 9. Payoffs and switching probabilities versus the malicious users' cost value

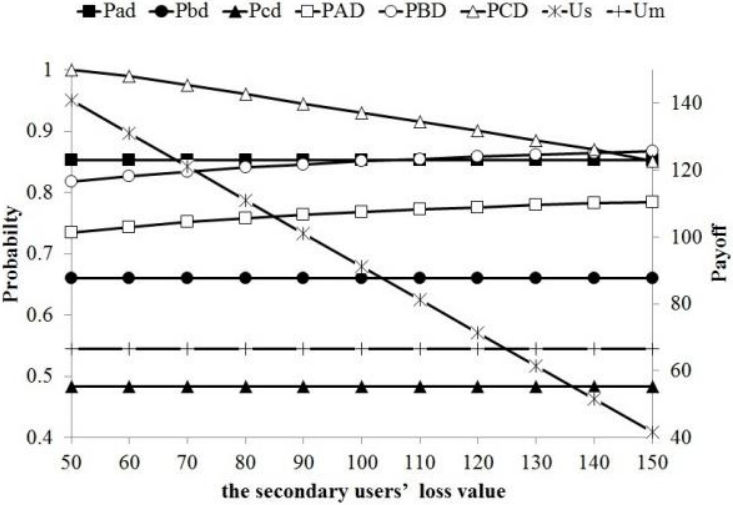

Fig. 10. Payoffs and switching probabilities versus the secondary users' lost value

The results of analytical simulation in Fig. 10 show that increasing $L_s$ does not change the switching probabilities of secondary users but decreases their payoff. Also, increasing $L_s$ does not change the payoff of malicious users but leads to change their switching probabilities.

Fig. 11 shows that by increasing $n_m$, the motivation of malicious users for switching to bands of state $D$ is reduced, because with increasing $n_m$, they become confident about their ability to attack more secondary users and thus will prefer to stay in their bands. For the same reasons, secondary users know that with increasing $n_m$, the probability of finding a free band is reduced. So, they prefer to stay in their states, but this reduction is not significant (Fig. 11). With increasing $n_m$, the payoff of malicious users is increased. The payoff of secondary users starts to decrease with increasing $n_m$ which finally leads to a negative payoff. In the negative payoff, secondary users must leave the network for the next time slot.

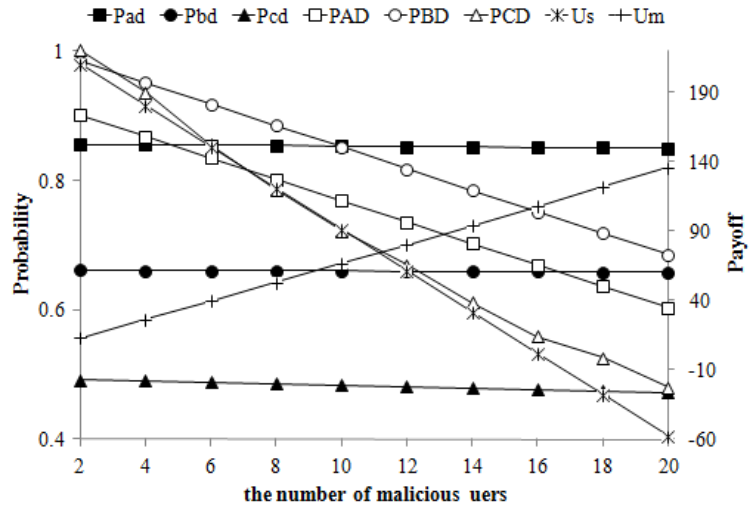

Fig. 11. Payoffs and switching probabilities versus number of the malicious users

## V. Conclusion

The radio frequency is an open medium, thus jamming attack can be a possible attack for the cognitive radio networks. It prevents the secondary users from accessing the spectrum band. However, malicious users can coordinate with each other to perform a coordinated jamming attack.

In this paper, the performance of cognitive radio networks was investigated upon a coordination jamming attack. In this model, both malicious and secondary users have a central base station and send the information of their bands to malicious-BS and secondary-BS, respectively. According to the information received from users, secondary-BS and malicious-BS distribute their users over appropriate bands. The spectrum band is classified into several states and according to the classification a game theoretical model is applied to analyze the behavior of both malicious and secondary users in the coordinated jamming attacks.

The values of cost in switching to other bands were defined for secondary and malicious users as well as the reward values for successful transmission of secondary users and successful attack of malicious users. Meanwhile, the total payoff was calculated for secondary and malicious users. Then, their optimal switching probabilities were computed using the game theory. With these optimal switching probabilities, secondary and malicious users can opt the most suitable bands against rival and prevent from wasting their energies. Moreover, effects of changing the values of cost and reward values for secondary and malicious users was also discussed on their total payoffs and switching probabilities.

In this contribution, equal occupation probabilities were proposed by primary users for all bands in the network. Future works can concentrate on studying a coordinated jamming attack in a spectrum band having different occupation probabilities by primary users.